# Tunable Antenna Coupled Intersubband Terahertz Detector

Nutan Gautam[1], Jonathan Kawamura[2], Nacer Chahat[2], Boris Karasik[2], Paolo Focardi[2], Samuel Gulkis[2],
Loren Pfeiffer[3], and Mark Sherwin[1]

[1]Institute of Terahertz Science and Technology, UC-Santa Barbara, Santa Barbara, CA, 93106 USA
[2]Jet Propulsion Laboratory, California Institute of Technology, Pasadena, CA, 91109 USA
[3]Department of Electrical Engineering, Princeton University, Princeton, NJ 08544 USA

*Abstract*— **We report on the development of a tunable antenna coupled intersubband terahertz (TACIT) detector based on GaAs/AlGaAs two dimensional electron gas. A successful device design and micro-fabrication process have been developed which maintain the high mobility ($1.1 \times 10^6$ cm$^2$/V-s at 10K) of a 2DEG channel in the presence of a highly conducting backgate. Gate voltage-controlled device resistance and direct THz sensing has been observed. The goal is to operate as a nearly quantum noise limited heterodyne sensor suitable for passively-cooled space platforms.**

## I. INTRODUCTION

THz mixers for astronomical applications are dominated by low-noise superconducting hot electron bolometers (HEB) operating at liquid Helium temperatures and Schottky diode mixers which can operate at ambient temperature but suffer from much higher noise. The latter are the only option for planetary missions where active cryocooling is not possible. The Schottky mixers however are not as low noise as HEB mixers and can hardly be used beyond 1 THz in view of the lack of sufficiently powerful solid-state oscillators. Here, we report on an alternative tunable antenna coupled intersubband terahertz (TACIT) [1] detector. The goal of this effort is to develop mixers for planetary applications with noise comparable to superconducting HEBs at temperatures accessible to passively-cooled space platforms (>50K), along with a gate bias tunable THz sensing.

The active region of the TACIT detector is a MBE-grown two-dimensional electron gas (2DEG) confined in a 40nm GaAs/AlGaAs quantum well designed for intersubband absorption near 2.5THz. This active region is a high mobility system whose resistance has strong temperature dependence for temperatures below 77K. The active region is sandwiched between a Schottky front gate and a buried back gate made of a highly doped GaAs layer. The front and back gates can tune the resonant frequency as well as the strength of the intersubband transition [2]. Also, they couple the THz signal to the active region via a twin-slot antenna. The source and drain are ohmic contacts to the 2DEG which sense the change in its resistance. When THz signal falls on the detector, electrons are excited to the next subband, leading to an increase in the temperature of the electron gas. Theoretically, the 2DEG has been approximated as a hot electron bolometric system [1] and in the presence of a local oscillator signal, the intermediate frequency (IF) signal can be sensed between the source and drain terminals. The schematic of THz detection and IF generation inside a TACIT detector is shown in Fig. 1(a), and Fig. 1(b) shows the microscopic image of a fabricated TACIT device.

The impedance of the TACIT detector's active region was calculated by treating the two-dimensional electron gas as a dipole sheet under the influence of an oscillating current between the front and the back gates [1]. A twin-slot planar antenna circuit was designed to provide an impedance match to the TACIT device and to couple THz radiation efficiently from the gates to the active region.

## II. RESULTS

The heart of the TACIT detector is a high mobility two-dimensional electron gas. A high mobility structure is known to have a strong temperature dependent resistance for temperatures lower than 77K. A 40nm quantum well with high mobility design along with a conducting back gate layer was grown at Princeton University. We successfully measured the mobility of $3.6 \times 10^6$ cm$^2$/V-s at 10K with the back gate but minimal processing. The main processing challenge was to attain high mobility after completing the processing of TACIT detector, which involves multiple dry etches and metal sequences. We measured a maximum mobility of $1.1 \times 10^6$ cm$^2$/V-s at 10K after complete fabrication process for the TACIT detector.

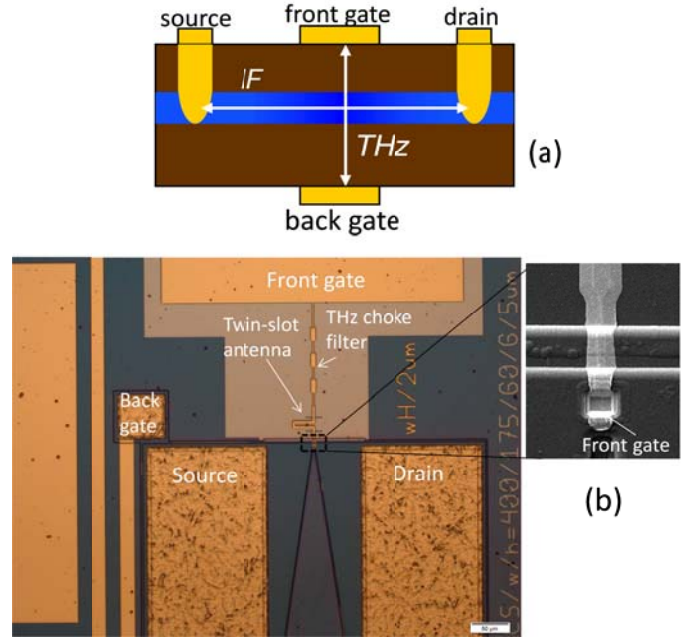

**Fig. 1** (a) Schematic of THz detection and intermediate frequency (IF) generation in a TACIT device, where 2DEG channel is represented by blue color. (b) Microscopic image of a fabricated TACIT device depicting source, drain, antenna, THz choke filter, front and back gates. The SEM micrograph of the active region (the square is the front gate) is shown in the inset.

The active region, shown in the inset of Fig. 1, has the dimension of 4μm×6μm. Dry plasma etching has been carried out to define the structure and to remove the conducting GaAs layer from the back plane metal region. Ni/AuGe/Ni/Au metal sequence was used to deposit ohmic metal for the source, drain, and backgate contacts and was subsequently annealed. The source and drain metal pads dimension is 396μm×167μm to minimize the contact resistance. A Ti/Pt metal sequence is used to deposit back plane metal to define a twin slot antenna and front Schottky gate on the active region. A silicon nitride-silicon dioxide-silicon nitride dielectric sequence is used to deposit a 1.2um thick dielectric layer on top of the back plane metal containing the antenna. A front gate biasing metal pad and microstrip containing a THz choke filter are deposited using Ti/Pt/Au on top of the dielectric layer and contact the front gate metal already defined on the device. The choke filter prevents coupling THz radiation from the antenna to the front gate contact pad.

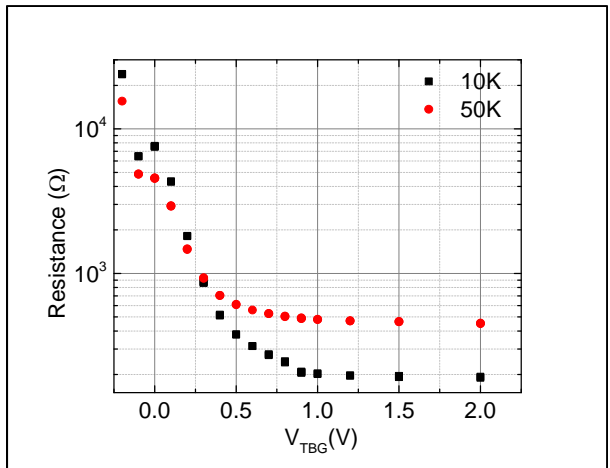

**Fig. 2** Source drain resistance of TACIT device as a function of gate bias applied between front gate and back gate at 10K and 50K.

We first carried out dc current-voltage measurements on the fabricated device. Figure 2 shows the device resistance as a function of gate bias at 10K, and 50K. The device resistance can be tuned from 10kΩ range down to few hundred ohms. This suggests that the front and the back gates function in the expected manner. The change in the device resistance is due to the change in the sheet charge density of the quantum well.

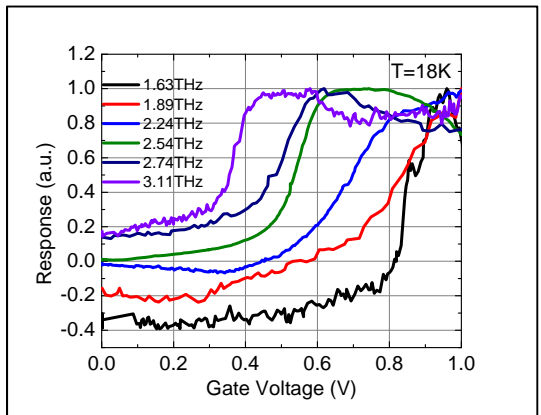

**Fig. 3** Tunable response of the TACIT detector has been demonstrated where the response of the detector can be tuned by tuning voltage bias across the front and the back gate.

We carried out direct detection measurements by mounting the device on a silicon lens, assembling the device into a receiver cryostat and illuminated it with a $CO_2$-pumped molecular gas far-infrared/THz laser. A photosignal was measured between the source and the drain terminals, while gate bias was tuned across the front and back gates. As shown in Fig. 3, for a given fixed THz pump frequency, the photoresponse is sharply enhanced above a threshold $V_T$. The value of $V_T$ depends strongly on the frequency of the THz pump. A key result of the study is this demonstration of voltage-tunable THz detection.

Figure 4 shows dependence of the threshold voltage $V_T$ on THz pump frequency. We define $V_T$ as the bias value at which the response reaches the average of its peak value and value close to zero applied bias. With increasing pump frequency, $V_T$ decreases approximately linearly. Further detailed characterization and theory are needed to understand the dependence of the photoresponse on THz pump frequency and voltage.

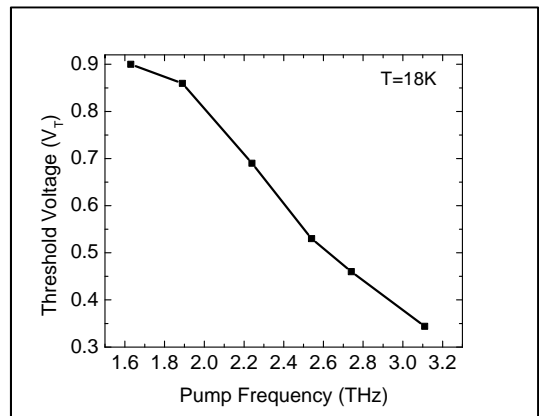

**Fig. 4** Dependence of threshold voltage $V_T$ on Terahertz pump frequency.

### III. SUMMARY

We have designed, fabricated, and demonstrated a tunable antenna coupled intersubband terahertz (TACIT) detector where THz response can be enhanced by the gate voltage applied between the front and back gates. For this, we developed a fabrication and design process to maintain the high mobility of a 2DEG channel which is the building block of the TACIT detector. Voltage-tunable THz photoresponse has been demonstrated for THz pump frequencies between 1.6 and 3.1 THz

### ACKNOWLEDGEMENTS

We would like to thank National Aeronautics and Space Administration for research funding. This research was done in part at the Jet Propulsion Laboratory, California Institute of Technology, under a contract with the National Aeronautics and Space Administration